# An Investor-Oriented Metric for the Art Market

**Ventura Charlin**
*V.C. Consultants*
*Los Leones 1300, Suite 1202*
*Santiago, CHILE*
e-mail: ventcusa@gmail.com

and

**Arturo Cifuentes**
*Financial Regulation Center*
*Faculty of Economics and Business*
*University of Chile*
*Santiago, CHILE*
e-mail: arturo.cifuentes@fen.uchile.cl

Ventura Charlin is an applied statistician with V.C. Consultants, a firm based in Santiago, Chile.

Arturo Cifuentes is a professor at the Faculty of Economics and Business, University of Chile, in Santiago, Chile, as well as the Academic Director of its Financial Regulation Center (CREM).

**Acknowledgements**

The authors thank Robert Yang whose technical expertise was instrumental in preparing the database. In addition, we would like to acknowledge Professor James MacGee (University of Western Ontario) for his valuable comments.

# An Investor-Oriented Metric for the Art Market

In the last thirty years, the art market −and especially, the market for paintings−has received an increasing amount of attention from economists and financial analysts. They have brought to this field many quantitative techniques already employed in more conventional markets. A topic that has received a great deal of attention is returns, specifically, how to compute return indices for the art market. This is a challenging task not only because this market is fairly illiquid, at least compared with equities and bonds, but also because of its heterogeneity: every painting is essentially a unique object.

Several authors have employed hedonic pricing models (HPMs) to estimate returns (Chanel, Gerard-Varet, and Ginsburgh [1994, 1996], de la Barre, Docclo, and Ginsburgh [1994], Edwards [2004], and Renneboog and Spaenjers [2011, 2013]). Such models are suitable to manage product variety and can use all the available data. Their drawback, however, is that their application is limited by the explicatory power of the variables selected and sometimes it is difficult to fit a good model to the data (the academic literature frequently reports models with values of $R^2$ around 60% or below).

A second alternative to estimate returns is to rely on repeat-sales regressions (Anderson [1974], Baumol [1986], and Goetzmann [1993]). While this approach has the advantage of using price data referring to the same object it has two disadvantages: a potential selection bias and the fact that it only employs a small subset of the available information, typically, below 25%.

Ginsburgh, Mei and Moses [2006] provide an excellent discussion on the merits of each approach plus a fairly complete literature review. Mei and Moses [2002], Renneboog and Spaenjers [2011], Higgs and Worthington [2005], Agnello and Pierce [1996], and artnet Analytics [2012] have dealt with the construction of art indices based on the two above-mentioned techniques or hybrid combinations of them.

Previous researchers have also focused on other topics. Just to name a few: Galenson [1999, 2000, 2001], Galenson and Weinberg [2000], and Ginsburgh and Weyers [2006] have looked at the creativity cycle of several artists (that is, the age at which they produced their best work). Renneboog and Van Houte [2002], Worthington and Higgs [2004], Renneboog and Spaenjers [2011], and Pesando [1993] have compared the returns of certain segments of the art market vis-à-vis more conventional investments. Coate and Fry [2012] have investigated the death-effect in the price of paintings. Edwards [2004] and Campos and Barbosa [2009] have looked at the performance of Latin American painters. Scorcu and Zanola [2011] used a hedonic model approach to study Picasso's paintings, while Campbell [2008] explored the diversification merits of investing in art. And, Sproule and Valsan [2006] questioned the accuracy of hedonic models compared with the appraisals of experts.

In summary, although a great deal has been learned about the financial aspects of the art market in recent years, much needs to be understood, especially, from the investor's perspective. Indices, although useful to describe broad market trends (*ex post*), are less useful to analyze the merits of individual paintings or to provide specific price guidance.

Therefore, our aim is to shift the focus towards the investor's viewpoint and move away from the purely econometric models, which, even though are interesting from an academic angle, offer little help to somebody concerned with making an investment decision. Thus, our goal is twofold: (i) to provide a new tool to enrich the investor' toolbox; and (ii) to facilitate the investors' decision-making process by making it easier to assess the merits of a specific painting using some simple quantitative analyses. The metric introduced herein achieves exactly that.

**A NEW FINANCIAL METRIC**

Paintings, notwithstanding their artistic qualities, are essentially two-dimensional objects that can command −sometimes− hefty prices. Based on this consideration, it makes sense to express the value of a painting not using its price but rather a price per unit of area (in this study, dollars per square centimeter). We call this figure of merit Artistic Power Value or APV. By normalizing the price, the APV metric intends to offer the investor a financial yardstick that goes beyond the price, while not attempting to control for the specifics of the painting beyond its area.

The intuitive appeal of this metric is obvious: simplicity, ease of computation, transparency, and straightforwardness. In fact, there is already a well-established precedent for this approach. For example, prices of other two-dimensional assets, such as raw land, are frequently quoted this way (e.g. dollars per acre, or euros per hectare). The same approach is sometimes used to quote prices of antique rugs.

More recently, many artisans, print makers, digital printing firms, and poster designers have started to quote price estimates using this same concept. Moreover, considering that the cost of materials (an important component of the production cost) employed in creating these two-dimensional objects is often estimated on a per-unit-of-area basis, the idea of extending the same notion to express the value of the final product is not far-fetched.

Finally, the rationale for using the APV metric is not to negate the individuality of each painting or to trivialize the artistic process. It is really an attempt to synthetize in one parameter the financial value of a painting (or artists or body of work) with the goal of making comparisons easier. Additionally, many APV-based computations (a point treated in more detail in the subsequent section) can offer useful guidance for pricing purposes.

Alternatively, we can think of the APV as an attempt to find a common factor to compare and contrast the economic value of otherwise dissimilar art objects. If we accept the thesis that two paintings −even if they are done by the same artist and depict a similar subject− are not only different but also unique, it is not possible to make a straight price-wise comparison. However, the APV metric, by virtue of removing the size-dependency, helps to make this comparison possible: in a sense the APV plays the role of unitary price.

**THE DATA**

We employ three data sets in this study:

a. Data set A consists of 1,818 observations of Pierre-Auguste Renoir's paintings auction prices and their characteristics covering the period [March 1985 – February 2013]. The database was built based on information provided by the artnet database (www.artnet.com).

b. Data set B consists of 441 observations of Henri Matisse's paintings auction prices and their characteristics covering the period [May 1960 – November 2012]. The database was built based on information provided by the artnet database (www.artnet.com) and was supplemented by additional auction data from the Blouin Artinfo website (www.artinfo.com).

c. Finally, data set C consists of 2,115 observations of paintings covering the period [March 1985 – February 2013]. This data set gathers information from six artists (Alfred Sisley, Camille Pissarro, Claude Monet, Odilon Redon, Paul Gauguin, and Paul Signac) and was based on auction information provided by the artnet database.

We adjusted all prices to January 2010 U.S. dollars (using the U.S. CPI index) and expressed them in terms of premium prices (we modified hammer prices and expressed them in terms of equivalent premium prices when appropriate). In addition, we eliminated all observations where the selling price was below US$ 10,000 or the APV was less than 1 US$/cm$^2$. Sotheby's and Christie's dominate the data sets, as together they account for 86% of the sales.

The selection of artists was somewhat arbitrary. The chief consideration was to effectively examine the merits of the APV metric without regard to the qualities of the painters selected.

Renoir was an ideal choice because of the high number of observations available, which were distributed over a long period of time, and without time gaps. This situation facilitates the comparison between the APV metric and the HPMs (which require many data points to be built). Matisse data had the advantage of being distributed over a longer time span, but included less observations, and had a few time-gaps. Data set C, despite its strong impressionist flavor, was not aimed at capturing in full the characteristics of the impressionist movement; it represents a group of painters who happened to live roughly at the same time and for which there were enough observations to make certain computations feasible. Nevertheless, and simply for convenience, in what follows we refer to this group as the Impressionists group. Renoir, despite his strong impressionist credentials was purposely left out of data set C. Otherwise, he would have dominated the group, making it highly correlated with data set A: an undesirable situation given the need to test the APV metric under different scenarios.

In summary, the selection of artists was not done with the idea of deriving any specific conclusion regarding these painters or the artistic tendencies they represented; the leading consideration was to showcase the attributes and benefits of the APV metric.

Exhibit 1 summarizes the key features of the three data sets. Exhibit 2 describes in more detail the characteristics of the painters in the Impressionist group (data set C). Notice that the APV distribution is far from normal: the differences between the arithmetic mean (average) values and the medians are manifest, with the means always higher than the medians. Additionally, the values of the skewness and kurtosis reveal a strong positively skewed distribution with fat tails. The Jarque-Bera (JB) statistic and its corresponding p-value (close to 0.000 for each of the three data sets) indicate that the APV is not normally distributed. These facts should serve as a warning against APV-based projections based on normality assumptions. Finally, the relatively high values of the coefficient of variation for several artists (Renoir and Matisse exhibit the most variability) are somehow evidence of what experts already know: even masters are uneven producers and their paintings differ greatly in quality. Whether ranking artists by their average or median APV values is consistent with the critics' assessment of their merits, it is a topic we leave for others to decide.

**EXHIBIT 1**
**Description of the three data sets and key statistics**

|  | Data Set: A | Data Set: B | Data Set: C |
|---|---|---|---|
| Artist | Pierre-Auguste Renoir | Henri Matisse | Impressionists group |
| Born–Died | 1841–1919 | 1869–1954 | NA |
| Number of Sales | 1,818 | 441 | 2,115 |
| Period of Sales | Mar 1985–Feb 2013 | May 1960–Nov 2012 | Mar 1985–Feb 2013 |
| Average APV (US$/cm$^2$) | 646 | 803 | 537 |
| Standard Deviation (US$/cm$^2$) | 1,331 | 1,332 | 786 |
| Coefficient of Variation | 2.06 | 1.66 | 1.46 |
| Median APV (US$/cm$^2$) | 377 | 308 | 311 |
| Skewness | 15.56 | 3.87 | 4.86 |
| Kurtosis | 344.06 | 19.83 | 31.87 |
| Jarque-Bera | 9,040,581.38 | 8,328.44 | 97,801.30 |
| JB *p*-value | 0.000 | 0.000 | 0.000 |

**EXHIBIT 2**
**Detailed characteristics and key statistics of the artists included in data set C**

| Artist | Number of Sales | Born–Died | Average APV (US$/cm$^2$) | Standard Deviation (US$/cm$^2$) | Coeff. of Variation | Median APV (US$/cm$^2$) |
|---|---|---|---|---|---|---|
| Alfred Sisley | 341 | 1839–1899 | 389 | 282 | 0.73 | 313 |
| Camille Pissarro | 586 | 1839–1903 | 432 | 335 | 0.78 | 338 |
| Claude Monet | 581 | 1840–1926 | 760 | 999 | 1.31 | 411 |
| Odilon Redon | 193 | 1840–1916 | 167 | 156 | 0.93 | 118 |
| Paul Gauguin | 167 | 1848–1903 | 1,138 | 1,631 | 1.43 | 465 |
| Paul Signac | 247 | 1863–1935 | 353 | 454 | 1.28 | 202 |

## APPLICATION OF THE APV METRIC

This section demonstrates the usefulness of the APV metric with the help of some examples.

**Comparisons among all artists**

The fact that the APV follows a highly non-normal distribution calls for comparisons to be based on the median rather than the average value. To this end we employ the median comparison test using the Price-Bonett variance estimation for medians (Price and Bonett [2001], and Bonett and Price [2002]), described in Wilcox's [2005] review of methods for comparing medians.

Exhibit 3 summarizes the results of such comparison. The median values for each artist are shown along the diagonal with the values decreasing from top-left to bottom-right: Matisse has the highest value (513 US\$/cm$^2$) while Redon the lowest (118 US\$/cm$^2$). The remaining entries in the table can be interpreted, using matrix notation, as follows: the (i, j) entry represents the median APV value of artist j minus the median APV value of artist i. Hence, Pissarro's median APV exceeds that of Signac by 136 US\$/cm$^2$ while there is no significant difference between Gauguin and Matisse's median APVs. These calculations, trivial by all accounts, offer a convenient way to rank artists. They also offer useful guidance for pricing purposes.

**EXHIBIT 3**
**Comparisons among the APV medians for all artists (1985-2012 sales only)**

| Median APV (diagonal) Difference between medians (off-diagonal) (US\$/cm$^2$) | Henri Matisse[a] | Paul Gauguin | Claude Monet | Pierre-Auguste Renoir | Camille Pissarro | Alfred Sisley | Paul Signac | Odilon Redon |
|---|---|---|---|---|---|---|---|---|
| Henri Matisse[a] | 513 | | | | | | | |
| Paul Gauguin | NS | 465 | | | | | | |
| Claude Monet | 102** | NS | 411 | | | | | |
| Pierre-Auguste Renoir | 136*** | 88* | 34* | 377 | | | | |
| Camille Pissarro | 175*** | 127** | 73*** | 39*** | 338 | | | |
| Alfred Sisley | 200*** | 152** | 98*** | 64*** | 25* | 313 | | |
| Paul Signac | 311*** | 263*** | 209*** | 175*** | 136*** | 111*** | 202 | |
| Odilon Redon | 395*** | 347*** | 293*** | 259*** | 220*** | 195*** | 84*** | 118 |

**NOTE:** [a]: Median calculated from sales from 1985-2012 only; NS: Not Significant; *p<.10; **p<0.05; ***p<0.01

**Portrait versus landscape orientation for a given artist**

Certain painters, Modigliani for instance (not part of this study) decidedly preferred the portrait (or vertical) orientation. Sisley and Signac, on the contrary, favored the landscape orientation. Exhibit 4 compares, for all the artists considered here, the median APV as a function of the orientation using the median-comparison algorithm already described. Note that in order to have similar periods for all comparisons among artists we only considered the sales between 1985 and 2012 for Matisse. The results are interesting and far from obvious. In the case of Sisley and Pissarro, the painting orientation does not affect the APV in a significant way. For Matisse and Renoir, the difference in median APV values is highly relevant indicating a market preference for their vertically orientated production. And, Monet and Signac were judged by the market as better at doing landscape-oriented paintings. In conclusion, the orientation of a painting, in most cases, has a definite influence on its market value.

**EXHIBIT 4**
**Comparisons of APV medians: portrait (vertical) versus landscape (horizontal) oriented paintings for each artist**

| Artist | Portrait | | Landscape | | Portrait versus Landscape Difference US$/cm$^2$ | P-Value |
| --- | --- | --- | --- | --- | --- | --- |
| | Number of Sales | Median APV (US$/cm$^2$) | Number of Sales | Median APV (US$/cm$^2$) | | |
| Alfred Sisley | 21 | 298 | 321 | 317 | -19 | NS |
| Camille Pissarro | 132 | 327 | 450 | 346 | -19 | NS |
| Claude Monet | 124 | 352 | 440 | 426 | -74 | <0.10 |
| Henri Matisse | 203 | 498 | 237 | 199 | 299 | 0.000 |
| Odilon Redon | 133 | 131 | 53 | 84 | 47 | <0.01 |
| Paul Gauguin | 81 | 580 | 86 | 328 | 252 | <0.05 |
| Paul Signac | 23 | 129 | 224 | 212 | -83 | <0.05 |
| Pierre-Auguste Renoir | 843 | 505 | 949 | 289 | 216 | 0.000 |

**NOTE:** Paintings with height=width are excluded from the table. NS: Not significant.

**Comparisons of different subjects for the same artist**

Exhibits 5, 6, and 7 display the median APV value, for each artist, as a function of three dummy variables, namely: (i) Still life; (ii) Landscapes and (iii) People (whether the painting shows one or several human figures regardless of the amount of detail); 0 refers to the absence of the condition.

**EXHIBIT 5**
**Comparisons of APV medians: still-life versus no-still-life for each artist**

| Artist | Subject: Still-Life=Yes | | Subject: Still-Life=No | | Difference US$/cm² | P-Value |
|---|---|---|---|---|---|---|
| | Number of Sales | Median APV (US$/cm²) | Number of Sales | Median APV (US$/cm²) | | |
| Alfred Sisley | NA | NA | NA | NA | NA | NA |
| Camille Pissarro | NA | NA | NA | NA | NA | NA |
| Claude Monet | 59 | 279 | 522 | 424 | -145 | <0.05 |
| Henri Matisse | 69 | 335 | 372 | 308 | 27 | NS |
| Odilon Redon | 58 | 214 | 135 | 86 | 129 | 0.000 |
| Paul Gauguin | 24 | 821 | 143 | 411 | 409 | <0.05 |
| Paul Signac | NA | NA | NA | NA | NA | NA |
| Pierre-Auguste Renoir | 364 | 302 | 1454 | 396 | -94 | 0.000 |

NA: Not enough sales for this artist in this subject (<10 sales). NS: Not significant.

**EXHIBIT 6**
**Comparisons of APV medians: Landscapes versus no-landscapes for each artist**

| Artist | Subject: Landscapes=Yes | | Subject: Landscapes=No | | Difference US$/cm² | P-Value |
|---|---|---|---|---|---|---|
| | Number of Sales | Median APV (US$/cm²) | Number of Sales | Median APV (US$/cm²) | | |
| Alfred Sisley | 282 | 311 | 59 | 321 | -10 | NS |
| Camille Pissarro | 325 | 342 | 261 | 340 | 2 | NS |
| Claude Monet | 410 | 424 | 171 | 355 | 69 | <0.10 |
| Henri Matisse | 143 | 161 | 298 | 459 | -298 | 0.000 |
| Odilon Redon | 42 | 61 | 151 | 135 | -74 | 0.000 |
| Paul Gauguin | 58 | 288 | 109 | 649 | -361 | 0.000 |
| Paul Signac | 103 | 218 | 144 | 200 | 18 | NS |
| Pierre-Auguste Renoir | 478 | 267 | 1340 | 429 | -162 | 0.000 |

NS: Not significant.

**EXHIBIT 7**
**Comparisons of APV medians: people (one or many persons) versus no-people for each artist**

| Artist | Subject: People=Yes | | Subject: People=No | | Difference | P-Value |
| --- | --- | --- | --- | --- | --- | --- |
| | Number of Sales | Median APV (US$/cm$^2$) | Number of Sales | Median APV (US$/cm$^2$) | US$/cm$^2$ | |
| Alfred Sisley | NA | NA | NA | NA | NA | NA |
| Camille Pissarro | 71 | 267 | 515 | 348 | -82 | <0.05 |
| Claude Monet | 12 | 338 | 469 | 415 | -77 | <0.10 |
| Henri Matisse | 190 | 586 | 251 | 206 | 381 | 0.000 |
| Odilon Redon | 25 | 56 | 168 | 124 | -67 | <0.01 |
| Paul Gauguin | 31 | 1115 | 136 | 388 | 727 | <0.01 |
| Paul Signac | NA | NA | NA | NA | NA | NA |
| Pierre-Auguste Renoir | 817 | 528 | 1001 | 285 | 243 | 0.000 |

NA: Not enough sales for this artist in this subject (<10 sales).

Clearly, certain artists are more appreciated for certain topics: Redon (see Exhibit 5) is more valued when executing still lives while the opposite happens with Renoir. Landscapes painted by Matisse, Gauguin, and Renoir (see Exhibit 6) are less desirable than other themes. And Gauguin, Renoir, and Matisse (see Exhibit 7) commanded higher prices when their paintings included people. These considerations are useful when appraising paintings.

**Life-cycle creativity patterns**

The idea behind this concept is to explore how the quality of an artist's paintings (using the APV metric as a proxy) evolves over time. That is, as a function of the age at which the painting was executed. Or more precisely, identify the period(s) at which the artist produced its most valuable work (financially speaking).

Exhibit 8 displays the median APV values, as a function of the age-at-execution for Renoir, Matisse, Monet, and Pissarro; i.e., the artists for whom we had more than 400 observations. The patterns shown are interesting as they reveal quite different tendencies. Renoir seems to have reached a peak around the mid-thirties and then experienced a slow decline. Matisse enjoyed a strong peak in his early forties, and a minor peak around his late fifties followed by a sequence of peaks and valleys in his late years. Monet's career is marked by two salient peaks: an early one (when he was thirty) and a later one (in his mid-sixties) while Pissarro's life is characterized by a more jagged curve that exhibited no significant decline in his old age and is more regular than those of either Monet and Matisse. This situation is somewhat consistent with the fact that Pissarro's coefficient of variation (0.78 from Exhibit 2) is lower than that of Monet (1.31) and Matisse (1.66).

**EXHIBIT 8**

**Life-cycle creativity curves**

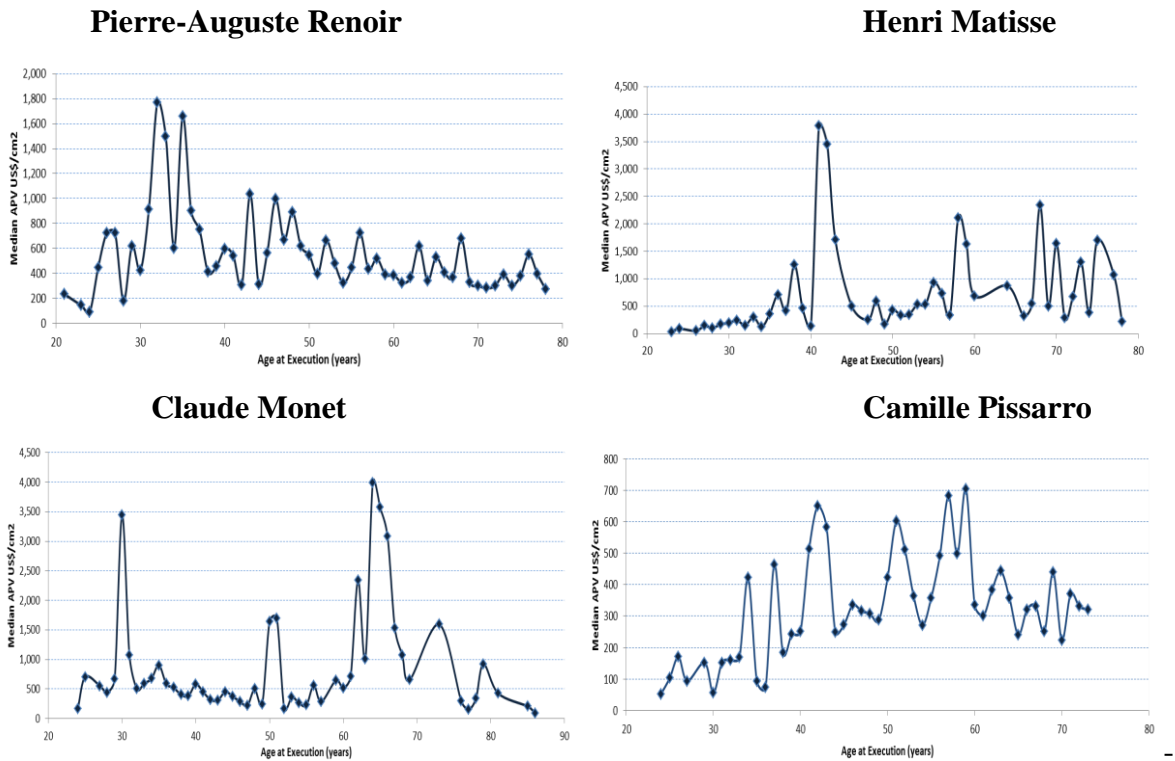

**Total returns for different artists or group of artists**

The total return computation is straightforward; first, we compute for each year, the average APV value (avg-APV). This is simply the sum of the APV values of all the paintings sold during the year divided by the total number of paintings sold. Then, the year-to-year total returns are computed based on the average APV values for two consecutive years. In short, the return between years $i$ and $i+1$ is simply $[\text{avg-APV}_{i+1}/\text{avg-APV}_i] - 1$.

We have purposely carried out this calculation using the average (mean) APV-value instead of the median. In general, it is customary to rely on the mean to estimate returns (regardless of the type of distribution) since the mean is better at capturing the influence of extreme values.

Leaving aside the ease of computation (undoubtedly an attractive feature) a valid question needs to be answered: What does this return mean?

The APV captures both, art market trends and supply-demand dynamics for the artist or artists considered, as it is based on actual sales. It does not intend to control the actual prices observed for any factor other than the area of the painting. Hence, the APV-based returns are really total (actual or realized) returns for the artist or artists in question (inflation has been removed since prices are expressed in January 2010 dollars).

**EXHIBIT 9**
**Year-to-year total returns (averages and standard deviations) and cumulative total return, using the APV and other relevant metrics**

| APV | Data Set A: Renoir | Data Set B: Matisse | Data Set C: Impressionists |
|---|---|---|---|
| Average Total Return | 8.16% | 43.78% | 8.30% |
| Standard Deviation Total Return | 35.58% | 172.19% | 31.47% |
| Cumulative Total Return* | 48.02% | 1195.65% | 184.21% |
| Average Year-APV (US\$/cm$^2$) | 593 | 661 | 501 |
| Standard Deviation Year-APV (US\$/cm$^2$) | 332 | 644 | 186 |
| Initial Year APV (US\$/cm$^2$)** | 360 | 70 | 227 |
| Final Year APV (US\$/cm$^2$)** | 533 | 902 | 645 |

*Cumulative total returns computed for 27 years for data sets A and C [1985-2012] and 52 years for data set B [1960-2012].

**Initial year-APV for data sets A and C is 1985 and for data set B is 1960. Final year-APV for all data sets is 2012.

Exhibit 9 summarizes the year-to-year return results: (i) average year-to-year total returns; and (ii) cumulative total returns for the relevant time-periods. The easiness with which one can compute these returns —contrasted, for example, with those estimated with HPMs— is striking. Exhibit 9 also shows the average APV over the years, their standard deviation, and the initial and final year-APV based on the year-sales considered for each data set.

**Repeat-sales vis-à-vis the entire (all-sales) data set**

Many analysts have estimated returns, for individual artists and groups of them, using only data from repeat sales. As pointed out before, a concern with this approach is that there could be a risk of selection bias. Exhibit 10 shows the median APV values for each of the artists considered using: (i) all the observations; and (ii) the repeat-sales subset. In two cases (Matisse and Renoir) the differences in medians are significant at the 5% level. And, in four of the remaining six cases the discrepancies are marginally significant (significant at the 10% level). We have performed the comparison between the two data sets using the median (instead of the mean) because of the marked non-normality of these samples.

**EXHIBIT 10**
**Comparisons of APV medians and total returns: all-sales versus repeat-sales for each artist**

| | All-sales | | | Repeat-sales | | |
|---|---|---|---|---|---|---|
| Artist | Number of Sales | Median APV (US$/cm$^2$) | Avg. Total Returns | Number of Sales | Median APV (US$/cm$^2$) | Avg. Total Returns |
| Alfred Sisley | 341 | 313 | 9.35% | 118 | 327 | 20.23% |
| Camille Pissarro | 586 | 338 | 6.41% | 146 | 378 | 13.69% |
| Claude Monet | 581 | 411 | 20.54% | 176 | 476 | 47.70% |
| Henri Matisse | 441 | 308 | 43.78% | 160 | 249 | 21.98% |
| Odilon Redon | 193 | 118 | 22.62% | 36 | 91 | 64.39% |
| Paul Gauguin | 167 | 465 | 57.98% | 37 | 612 | 115.64% |
| Paul Signac | 247 | 202 | 29.28% | 90 | 180 | 27.18% |
| Pierre-Auguste Renoir | 1,818 | 377 | 8.16% | 426 | 425 | 33.18% |

Finally, and somehow expectedly, the estimated returns (based on year-average APV-values) are quite different for the two groups (all-sales versus repeat-sales). The fact that in most cases the returns are higher when computed based on the repeat-sales set gives credibility to the hypothesis that paintings are more likely to be sold if they have increased in value.

These findings support the view that a selection bias cannot be ruled out when dealing with repeat-sales data. Thus, return estimates based on repeat-sales regressions (despite the claim that one has controlled for all the relevant factors) should be regarded with suspicion. The same goes for any other estimate based on repeat-sales information.

In conclusion, the examples discussed in this section show that the APV metric is a useful tool that can provide potential investors with a great deal of insight regarding the merits of an artist, groups of artists, or a particular painting.

**VALIDATION OF THE APV METRIC**

A useful way to assess the validity of the new metric is to compare the returns obtained with the APV and those obtained with the commonly used hedonic models.

In order to determine a fair yardstick for comparison purposes we carry out two steps. First, we estimate individual HPMs for each of the three cases (Renoir, Matisse, and the Impressionists) using the entire corresponding data set. And second, in each case, we evaluate the resulting HPM, for each year, using the average characteristics of the paintings sold during the year, to arrive at a representative price corresponding to each year, $P_i$ (where $i$ denotes a year index). The year-to-year HPM-based returns are computed based on these prices, using the expression $(P_{i+1}/P_i) - 1$. Thus, the idea is to use the HPM to estimate the total return.

The HPMs employ the natural logarithm of the painting selling price as the dependent variable. The independent variables (right-hand side of the regression equation) involve:

(i)    linear and higher-order polynomial expressions based on the age of the artist at the time the painting was executed;

(ii) in the case of data set C a dummy (binary) variable to account for the identity of the painter;

(iii) linear and higher-order polynomial expressions based on variables associated with the geometry of the painting (area, height, width, aspect ratio, and diagonal) plus binary dummy variables accounting for medium (canvas) and special topics (nudes, still lives, flowers, etc.); and

(iv) a sequence of dummy (binary) variables associated with the year the painting was sold.

The corresponding adjusted $R^2$'s (Renoir, Matisse, and Impressionists) are as follows: 0.75 (F= 137.47, $p<.0001$), 0.72 (F=18.78, $p<.0001$), and 0.67 (F= 77.39, $p<.0001$) respectively. In addition, we used White's (1980) test for heteroscedasticity and the null hypothesis of homoscedasticity in the least-squares residuals was not rejected in each of the three samples (results can be provided upon request).

Exhibit 11 shows the comparison between the average year-to-year total return estimated with (i) the APV metric; and (ii) the HPMs applied as described before. Both estimates, in all three cases, are in close agreement. This fact is also consistent with the high correlation values reported as well as the visual comparison presented in Exhibits 12, 13, and 14.

**EXHIBIT 11**
**Year-to-year returns: averages, standard deviations, and correlations (APV and HPM)**

|  | Data Set A: Renoir | Data Set B: Matisse | Data Set C: Impressionists |
|---|---|---|---|
| Average Total Return (APV) | 8.16% | 43.78% | 8.30% |
| Standard Deviation Total Return (APV) | 35.58% | 172.19% | 31.47% |
| Average Total Return (HPM) | 7.64% | 48.74% | 8.36% |
| Standard Deviation Total Return (HPM) | 38.12% | 162.67% | 33.35% |
| Correlation Total Ret. APV - Total Ret. HPM | 0.79 | 0.89 | 0.82 |
| Average Market Return (HPM) | 4.99% | 18.60% | 7.51% |
| Standard Deviation Market Return (HPM) | 24.16% | 67.64% | 27.67% |
| Correlation Total Ret. APV - Market Ret. HPM | 0.80 | 0.66 | 0.88 |

**EXHIBIT 12**
**Year-to-year total (APV and HPM) returns for Pierre-Auguste Renoir sales.**

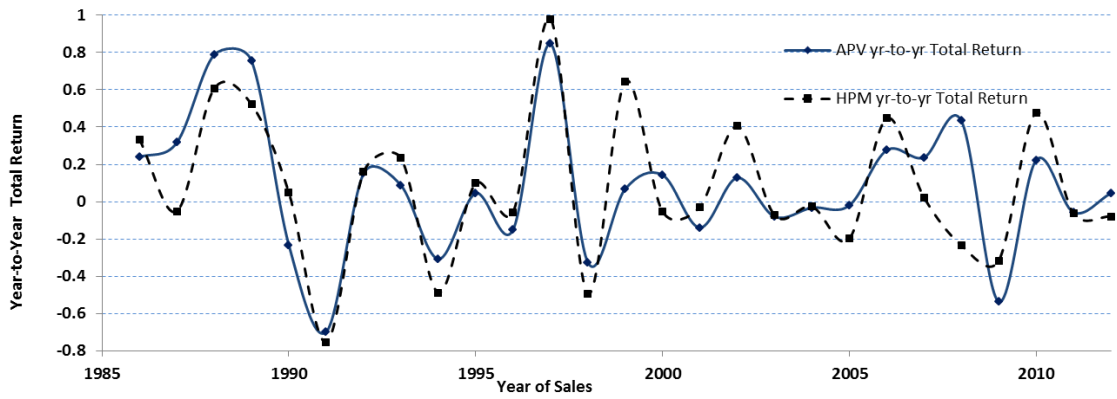

**EXHIBIT 13**
Year-to-year total (APV and HPM) returns for Henri Matisse sales.

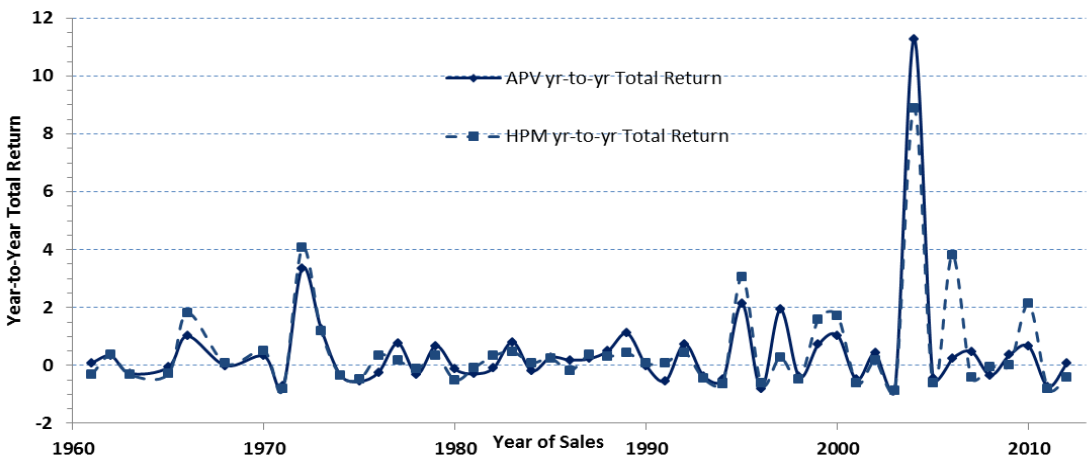

**EXHIBIT 14**
Year-to-year total (APV and HPM) returns for Impressionists group sales.

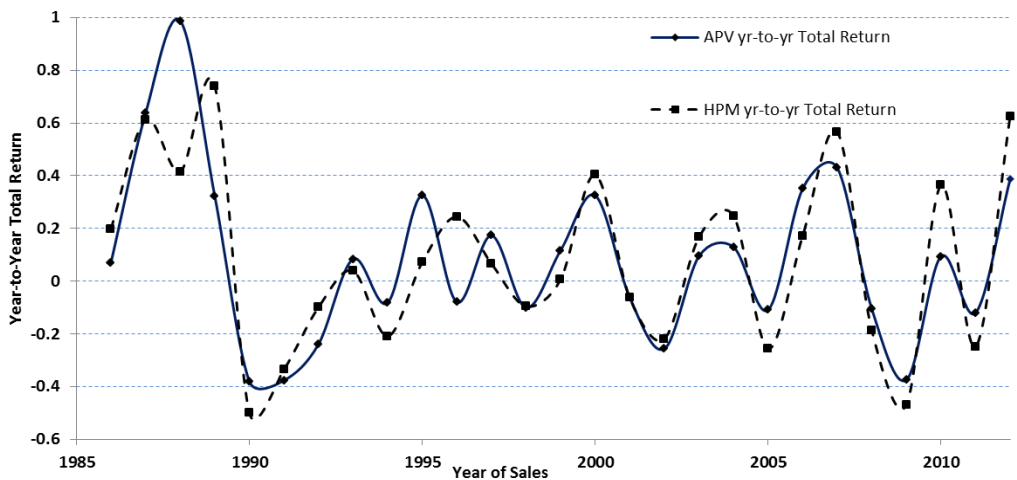

It might not seem evident that this is a fair comparison. However, we should notice that by evaluating the HPMs for each year, with the typical features of the paintings sold that year, one is capturing the two effects that influence returns: the market trend (reflected in the HPM coefficients associated with the time-dummy variables) and the specific characteristics of the paintings sold on a given year. The APV metric blends these two factors (market trends and paintings features) in one number. Therefore, the hedonic model framework (applied in the modified manner just described) seems appropriate to double-check the validity of the returns based on the APV metric. The conventional HPM-based returns estimated with the time-dummy coefficients would have captured only market trends and therefore would have been unsuitable for our comparison.

The fact that the APV-based returns, in all cases considered, yielded very similar results to those obtained with the hedonic models provides good evidence that the APV metric, despite its simplicity, offers results consistent with more conventional methods.

This high degree of consistency might seem surprising. However, the following two observations can explain, appealing partly to intuition, the success of the APV: (1) regressing the logarithm of the price on just the area of the painting, for the case of Renoir, Matisse, and the Impressionists, we obtained adjusted $R^2$'s values equal to 0.37, 0.26, and 0.33 respectively. Recall that the $R^2$'s values of the corresponding hedonic models were 0.75, 0.72, and 0.67 respectively. Hence, the APV metric −for all its roughness and simplicity− was able to explain, just by itself, almost half of what all the factors of the HPMs did; and (2) if we compute the correlation between the area of the paintings and the logarithm of the price for all the artists considered (Sisley, Pissarro, Monet, Matisse, Redon, Gauguin, Signac, and Renoir) we obtain the following (fairly high) values: 0.36; 0.65; 0.48; 0.51; 0.55; 0.59; 0.68, and 0.61 respectively. These observations provide some basis for making an argument that using the area of a painting as a normalization factor is not that eccentric or bizarre; it has some sound foundation.

We can also use the HPMs just described to estimate returns using a conventional approach, that is, by means of the time-dummy coefficients. These returns, as already explained, would only capture market trends and thus we refer to them as market returns as they do not capture the specific characteristics of the paintings under consideration explicitly. These returns are also shown in Exhibit 11. Even though they are highly correlated with the APV-based returns, they differ in absolute value, although less for the Impressionists. This should not come as a surprise as these returns, by definition, are not intended to account for the influence on the returns of the year-to-year variation in the characteristics of the paintings. Therefore, we claim that these market returns might be useful to detect −a posteriori− broad market tendencies, but they are not very illuminating when trying to price a painting or make comparisons.

**FUTURE APPLICATIONS**

The market for paintings lacks a widely accepted index or indices that could be used to design derivatives contracts for hedging and/or speculative purposes. We reckon that the reason is that the most popular indices (Mei-Moses index, artnet.com family of indices, AMR indices, etc.) while effective for the purpose they were designed −namely, tracking broad market trends− are unsuitable for financial contracts. The reason is that they involve several elements (proprietary databases, discretionary rules in terms of which sales should be included, ad hoc combinations of repeat-sales techniques coupled with some undesirable features of HPMs) that make them opaque and −at least in theory− vulnerable to manipulation. In contrast, indices such as the S&P 500 or the Barclays Capital bond indices family −which are based on well-defined and transparent rules− are easy to reproduce and difficult to game. Not surprisingly, derivatives contracts based on these indices have enjoyed wide market acceptance.

We think that the APV metric provides a natural tool to create well-defined indices that could be the foundation for a derivatives art market. If one wishes to design an index to represent a specific market segment −for example, the Impressionists− the main point is to agree on the artists that should be part of the index (a rule that must stay unaltered over time). Once this issue is settled what remains to agree upon is simply a mechanistic recipe to calculate the value of the index. For instance, it could be the average APV value of all the paintings sold in public auctions in the last twelve months as long as their values exceeded US$ 50,000.

A contract built around an index of this type could be used to gain exposure to this market or short it, in amounts much smaller than the typical price paid for a masterpiece. In that sense, these types of contracts could help to expand the investor base, and contribute to improve market liquidity. The operational details are similar, for instance, to those encountered in the agricultural derivatives market or commodities markets. This topic is presently under investigation by the authors.

**SUMMARY AND CONCLUSIONS**

We have introduced an easy-to-compute financial metric suitable for two-dimensional art objects that is both intuitive and transparent. It has several appealing features: it is difficult to game since not much discretion comes into its evaluation (unlike hedonic models that are data intensive and often exhibit lack of stability); it can be applied to artists for whom there are few observations, albeit with all the caveats appropriate for small data sets; it facilitates comparisons between artists, between different types of paintings by the same artist, or, paintings done by the same artist at different life-periods; it is also appropriate to explore artists' consistency, by looking at its standard deviation or coefficient of variation; and, finally, it can be employed to construct well-defined total-return indices to create financial derivatives.

However, it must be emphasized that the main goal of this new metric is to offer an investor a useful yardstick that captures, after normalizing by the area, a representative price. It is not the aim of the APV to control prices for other characteristics or to build a market index based on a time-independent ideal painting. For these reasons the APV metric is ideally suited to compute actual returns.

In terms of estimating returns, the APV metric offers three attractive features: (i) unlike repeat-sales regression models, it uses all the available data; (ii) unlike HPMs, whose effectiveness can depend substantially on the variables chosen and the analyst's skill to select them, the APV gives a unique value: the actual total return; and (iii) APV-based returns can always be computed regardless of the number of observations. On the other hand, HPM-based returns can be computed only in the limited number of cases where one has enough data, with the caveat that the accuracy of such return estimates is weakened by the explicatory power of the relevant model since the $R^2$ is never 1.

Some academics might feel that APV-based returns are contaminated, since we do not control for factors such as the type of painting (subject matter), geometric features beyond the area, and the host of other variables that hedonic models normally employ to explain the price (dependent variable). This argument might sound reasonable until we realize that when computing returns for stocks —for example— we never control for "other" factors such as revenue composition, number of employees, or sales volume. We simply record the change in stock price. Well, the same goes for APV-based returns. In fact, we argue that potential investors probably desire a metric that actually captures the revenue composition variation. In summary, the fact that APV-based returns do not control for any factors beyond the area rather than being a weakness of the metric is a source of strength.

Although the aim of this paper has been to incorporate a new tool to enhance the analyst's toolkit, an uncomfortable fact must be dealt with. Our findings undermine a great deal of previous research on this field. Specifically:

- We have provided evidence to be concerned about the reliability of returns computed with repeat-sales regressions. The fact that in other fields (real estate for example) repeat-sales indices such as the Case-Shiller index have gained wide acceptance is not relevant. In the U.S., real estate market repeat-sales account for more than 90% of sales. In the art market, repeat sales, depending on the segment, account at most for 20 or 25% of total sales. This fact, coupled with the striking differences between the characteristics of the repeat-sales subsets and the entire data set, casts doubt on the relevance of indices such as the Mei-Moses index.

- The fact that with a trivial computation we have been able to reproduce the returns estimated with the more numerically-intensive HPMs calls into question the usefulness of such approach. Moreover, the fact the HPMs cannot be used with sparse data (a common situation, especially for individual artists) undermines even further their potential benefits. And finally, we must note that returns estimated with HPMs, are just that: estimates. The explicatory power of such models is often weakened by $R^2$ coefficients that are frequently far smaller than 1; this situation makes HPMs even less attractive as tools to estimate returns.

- Several authors have investigated the validity of the CAPM model within the context of the art market. Although the results have been mixed we also think they have been irrelevant. The reason is that most authors —erroneously in our view— have placed on the left-hand side of the CAPM equation estimates of returns obtained, in general, via the time-dummy coefficients of a suitable hedonic model. We reckon that the correct approach is to place on the left-hand side of the CAPM equation actual total returns (which can be easily obtained with the APV metric) instead of estimates of market returns. This conceptual error calls to revisit most previous findings regarding this topic.

Finally, we hope investors, financial analysts, and researchers will be able to explore −and exploit− the merits of the APV metric in the near future. Our goal is simply to introduce a new tool, showcase a few applications, and perform some validation tests.

In summary, the main advantage of the APV is that it is a financial metric and not a modeling technique; therefore, it is what it is, and it can always be computed. In short, it can be useful or useless, but never wrong.